\documentclass{article}
\usepackage{graphicx} 
\usepackage{amsmath}
\usepackage{amssymb}
\usepackage{johd}
\usepackage{ulem} 
\usepackage{color}
\usepackage[dvipsnames]{xcolor}

\usepackage{cancel}
\usepackage[font=small,labelfont=bf]{caption, subfig}
\usepackage[margin=1in]{geometry}

\definecolor{custom}{HTML}{B8892D}

\newcommand\authormark[1]{\textsuperscript{#1}}
\title{Feedback-Controlled Self-Trapping of Solitons}

\author{
\begin{minipage}{0.95\textwidth}
\centering
Raonaqul Islam\authormark{1,*}, Pradyoth Shandilya\authormark{1}, Yunxiang Song\authormark{2},
Alioune Niang\authormark{1}, Gary Carter\authormark{1}, Thomas Carruthers\authormark{1},
Curtis Menyuk\authormark{1}, and Ergun Simsek\authormark{1}\\[3pt]
\footnotesize
\authormark{1}Department of CSEE, University of Maryland Baltimore County, Baltimore, MD 21250, USA\\
\authormark{2}John A. Paulson School of Engineering and Applied Sciences, Harvard University, Cambridge, MA 02138, USA\\[2pt]
\authormark{*}\texttt{raonaqul.islam@umbc.edu}
\end{minipage}}
\date{}

\begin{document}
\maketitle
\begin{abstract}
    \noindent Chip-scale microresonator-based frequency combs, commonly referred to as microcombs, are promising sources for field-deployable metrology applications. However, intrinsic microresonator noise can make it difficult to generate microcombs with sufficiently low noise for applications. A widely used approach to overcome this limitation is to trap the DKS in the microresonator by phase-modulating the pump laser, although integrating a low-noise modulation reference in a fully on-chip platform remains challenging. Here, we propose an optoelectronic self-trapping scheme in which the impact of intrinsic microresonator noise on microcomb repetition-rate jitter is suppressed through feedback-controlled phase modulation of the driving laser so that the intrinsic noise of the laser sets the ultimate performance limit. This modification eliminates the need for an external low-phase-noise RF source, thereby enabling chip integration. We develop both analytical and numerical models for the proposed system and benchmark its performance against conventional externally referenced bulk implementations, showing that the proposed approach achieves comparable microcomb noise suppression while maintaining chip-scale compatibility.
\end{abstract}

\section{Introduction}
Low-noise optical frequency combs are useful for time and frequency measurements \cite{Diddams:2020} because they can be utilized as a tool to connect the microwave and optical domains for optical atomic clocks \cite{Newman:2019,Oelker:2019}, time and frequency transfer \cite{Giorgetta:2013}, frequency synthesizers \cite{Spencer:2018}, and microwave oscillators \cite{Sun:2025}. Optical frequency combs generated by dissipative Kerr solitons (DKSs) in microresonators, known as soliton microcombs, have gained popularity due to their broad spectral bandwidth, low-power operation, SWaP (size, weight, and power) and chip integration compatibility \cite{Kippenberg:2018}. However, the repetition-rate ($f_{\text{rep}}$) of chip-scale microcombs suffers from non-negligible phase-noise dominated by thermal noise \cite{Drake:2020, Stone:2020, Huang:2019}, Raman soliton self-frequency shift (SSFS) related transduction noise \cite{Triscari:2023}, and noise induced by dispersive wave (DW) recoil \cite{Yang:2021}. Developing low-noise and chip compatible microcombs is a is a rapidly developing area of research due to recent advances in integrated photonics \cite{Zhu:21}. 

Previously, Jost et al.\ demonstrated that microcombs can be stabilized using active feedback to control the thermal drift in the noise source \cite{Jost:2015}. Their setup only works for large resonators, where $f_{\text{rep}}$ is not strongly affected by intrinsic noise, thereby making it unsuitable for chip integration. Weng et al.\ reported that the DKS inside the microresonator can be trapped by phase modulating the pump, resulting in the purification of the generated microwave signal \cite{Weng:2019}. Although \cite{Weng:2019} demonstrated substantial suppression of the overall noise inherited from a high-$Q$ ($1.3\times10^9$) 14.09-GHz microresonator, the use of a commercial microwave oscillator, a stable fiber laser, and an acousto-optic modulator limited the chip-scale compatibility of the system. Meanwhile, Moille et al. presented a technique to trap the soliton using Kerr-induced synchronization (KIS) \cite{Moille:2025}, but the system requires an additional tunable reference pump that is not easy to integrate on chip \cite{Guo:2019}. Recently, a fully passive opto-electronic oscillator (OEO) microcomb has been reported \cite{Long:2025}, which generates a 10.7-GHz platicon microcomb while simultaneously synthesizing a low-phase-noise microwave signal. However, the microwave phase-noise spectra with and without platicon formation are nearly identical, indicating that the intrinsic noise in this large mode volume microresonator is negligible. Meanwhile, Li et al. numerically show that the comb line at the location of a dispersive wave can be re-injected into the microresonator to lock $f_{\text{rep}}$ \cite{Li:2025} via KIS, where the re-injected dispersive wave acts as the reference pump. Such microresonators require highly accurate dispersion engineering, as being off by even a single mode number causes the filter to select the wrong comb line, leaving insufficient power for KIS. 

In small mode volume microresonators, the impact of intrinsic noise on $f_{\text{rep}}$ is greater due to timescale shortening of thermodynamic fluctuations \cite{Stone:2025}. In this work, we show that the microcomb repetition rate noise in small mode volume microresonators resulting from intrinsic noise sources can be reduced by using the microwave signal generated by sending the microresonator output to a high-speed photodetector to phase modulate the pump. Our approach traps the soliton and suppresses the impact of the intrinsic noise of the microresonator on the soliton repetition rate without the necessity of a bulky low-noise RF source. Realizing such a stabilization scheme is particularly important for low mode volume microcomb systems operating at high repetition-rates ($\sim100\; \text{GHz}$) \cite{Song:25:soliton, Song:26:kerr}, where the required optoelectronic components are becoming available on the same chip-scale platform \cite{Xu:22, Guo:22}.

Here, we discuss the fundamental noise limits of the proposed system. We consider a 100-GHz-FSR Si\textsubscript3N\textsubscript4 microring resonator with purely second-order dispersion. When the microresonator has only even-order dispersion terms, pump-detuning noise has a negligible effect on the repetition rate of the DKS \cite{Cherenkov:2017}. In practice, however, residual higher-order dispersion terms cause pump noise to transduce into the repetition rate via DW recoil, while the Raman SSFS transduces pump noise into the repetition rate even when dispersion is purely second order. As a result, the system's performance limit is set by pump-laser noise transduced through these pathways \cite{Liang:2015, Lucas:2020}. Low-noise lasers stabilized using stable coil resonators and isolated from environmental noise sources have recently been reported and are well suited to mitigating this pump-noise-limited regime \cite{Sun:2025:OFD, Liu:2022}. Moreover, the effects of Raman SSFS and DW recoil can themselves be suppressed using quiet-point engineering \cite{Triscari:2023}, which shifts the performance limit to the microresonator's thermorefractive noise at low offset frequencies and to the photodetector's shot noise at high offset frequencies \cite{Yi:2017}. Therefore, in Sec.~\ref{sec:results}, we first analyze the impact of a generic white noise source that is assumed to directly affect the DKS repetition rate, and then study the impact of thermorefractive noise that affects the resonance frequencies of the cavity.

\begin{figure}[!tb]
    \centering
    \includegraphics[width=1.0\linewidth]{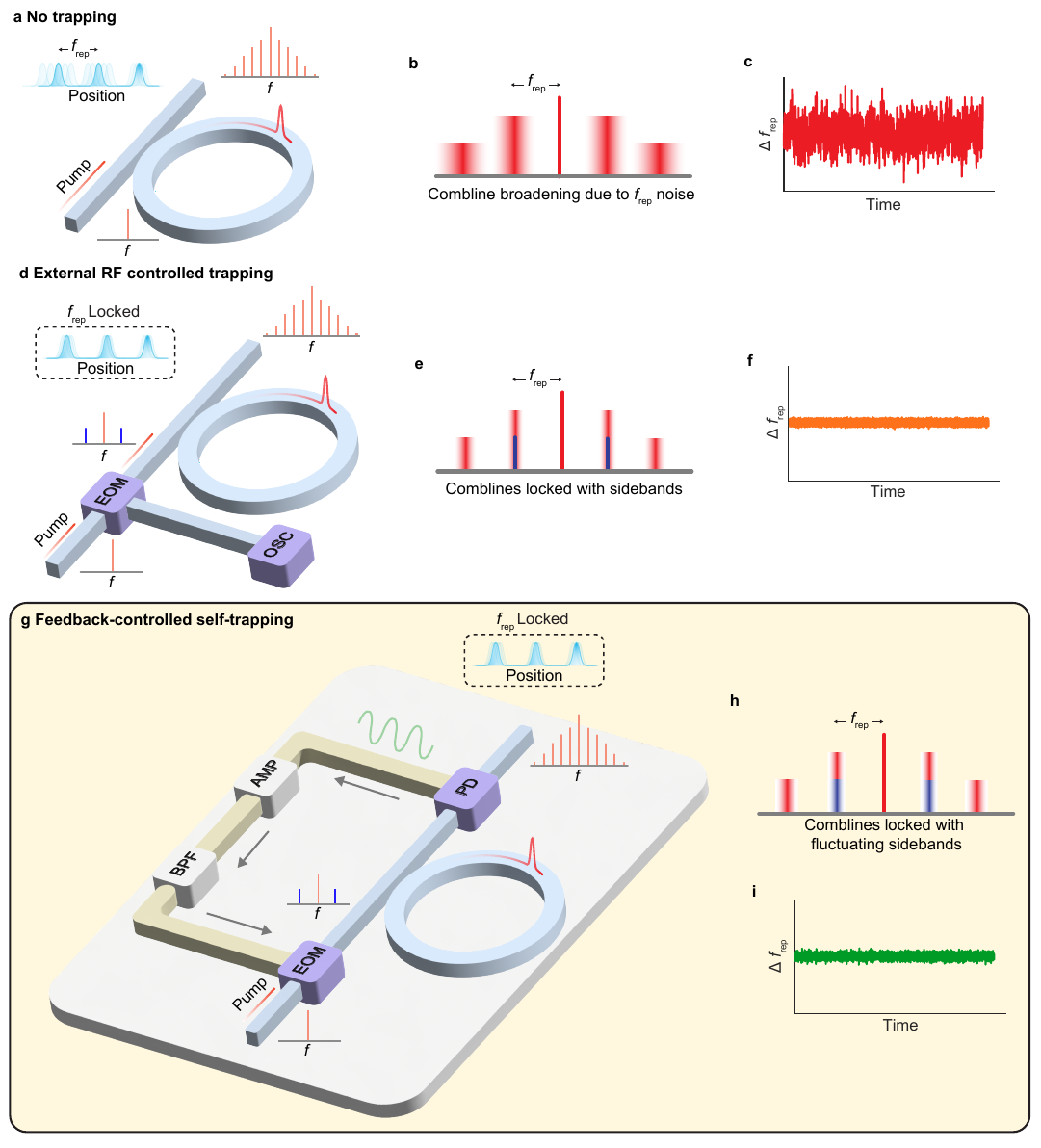}
    \caption{(a) DKS-microcomb generation using a microresonator pumped by a CW laser. (b) Comb lines of the resulting microcomb are broadened because of fluctuations in $f_{\text{rep}}$ due to noise (c). (d) DKS is trapped using phase modulation of the pump laser, resulting in reduced broadening of comblines (e) and a lower range of fluctuations in $f_{\text{rep}}$ (f). (g) Proposed self-trapping scheme, where trapping is achieved using feedback, consequently exhibiting less broadened comblines (h) and a decrease in $f_{\text{rep}}$ fluctuations (i). BPF: Band pass filter, AMP: RF amplifier, EOM: electro-optics modulator, PD: photodetector, OSC: RF oscillator.}
    \label{fig:schematic}
\end{figure}

\section{Concept of Self-Trapping}
When a Kerr-nonlinear microresonator with anomalous group-velocity dispersion is driven by a continuous wave (CW) laser, the interplay between dispersion, Kerr nonlinearity, cavity dissipation, and parametric gain enables the formation of dissipative Kerr soliton (DKS) states \cite{Kippenberg:2018}. The periodic extraction of these intracavity pulses produces an optical frequency comb with equally spaced spectral lines, which is known as a DKS microcomb, as illustrated in Fig.~\ref{fig:schematic}(a). The pulse-to-pulse spacing defines the repetition rate, $f_{\text{rep}}$, which closely follows the cavity free spectral range. However, intrinsic microresonator noise perturbs the group velocity of the soliton leading to random fluctuations of $f_{\text{rep}}$. (Fig.~\ref{fig:schematic}(b),(c)).

A well-established method to suppress these $f_{\text{rep}}$ fluctuations is to modulate the pump phase at a frequency near $f_{\text{rep}}$ using a low phase-noise RF reference. The modulation generates pump sidebands separated from the carrier by approximately $f_{\text{rep}}$ (Fig.~\ref{fig:schematic}(d)), which breaks the continuous translational symmetry of the soliton dynamics and traps the soliton \cite{Weng:2019,Taheri:2015, Moille:2025}. This trapping suppresses the fluctuations of $f_{\text{rep}}$ (Fig.~\ref{fig:schematic}(e),(f)). However, the requirement of an external low-noise RF source limits chip-scale integration.

Instead of using a bulky RF source, the microcomb repetition-rate beat note can be directly obtained by photodetection of the comb, yielding an RF signal at the frequency of $f_{\text{rep}}$ (Fig.~\ref{fig:schematic}(g)). This signal can then be amplified and spectrally filtered to drive an electro-optic modulator (EOM), thereby generating pump sidebands in a self-referenced manner, analogous to the externally referenced scheme. In this feedback-based approach, the generated sidebands inherit fluctuations correlated with the microcomb $f_{\text{rep}}$ jitter which may affect the overall performance of the system. Nevertheless, we will show here that the resulting modulation-induced trapping potential remains sufficient to trap the soliton, leading to improved stabilization of $f_{\text{rep}}$ (Fig.~\ref{fig:schematic}(h),(i)). In the following sections, we first describe the numerical modeling approach and then derive analytical expressions for the noise reduction. Finally, we present the power spectral density (PSD) of $f_{\text{rep}}$ fluctuations for the untrapped, externally trapped, and self-trapped configurations.

\section{Numerical and Theoretical Analysis}
To obtain numerical DKS solutions, we employ the Lugiato--Lefever equation (LLE) \cite{Chembo:2013}, expressed as

\begin{equation}
    \frac{\partial \mathcal{A}}{\partial \tau}= -\left(\frac{\kappa}{2}+i\sigma\right)\mathcal{A}-\frac{\zeta_2}{2}\frac{\partial^2\mathcal{A}}{\partial\theta^2}-ig_0|\mathcal{A}|^2\mathcal{A}+\frac{\kappa}{2}\mathcal{F},
    \label{eq_1}
\end{equation}

\noindent where $\mathcal{A}$ denotes the slowly varying intracavity field envelope, $\kappa$ is the cavity decay rate, $\sigma$ is the detuning between the pump laser and the nearest cavity resonance, $\zeta_2$ is the second-order dispersion coefficient, $g_0$ is the four-wave mixing gain, and $\mathcal{F}$ is the pump amplitude. We normalize time in Eq.~\ref{eq_1} to photon lifetime and obtain the LLE in dimensionless form as \cite{Taheri:2015,Chembo:2013}

\begin{equation}
    \frac{\partial \psi}{\partial t}=-(1+i\alpha)\psi-i\frac{\beta_2}{2}\frac{\partial^2\psi}{\partial\theta^2}+i|\psi|^2\psi+F_0,
    \label{eq_2}
\end{equation}

\noindent where $\psi=\mathcal{A}/\mathcal{A}^*_{\mathrm{th}}$, $|\mathcal{A}_{\mathrm{th}}|=(\Delta\omega_0/2g_0)^{1/2}$, $F_0=\mathcal{F}/|\mathcal{A}^*_{\mathrm{th}}|$, $t=2\tau/\kappa$ is the normalized time, $\beta_2=2\zeta_2/\kappa$ is the normalized second-order dispersion parameter, and $\alpha$ is the normalized detuning.

The linear stability of the DKS solution is evaluated by introducing a small perturbation and tracking its evolution using dynamical methods \cite{Qi:19}. The steady-state solution of Eq.~\ref{eq_2} is obtained using the Levenberg--Marquardt algorithm \cite{Levenberg:1944} and is denoted by $\psi_0$. Linearizing Eq.~\ref{eq_2} about $\psi_0$ yields

\begin{equation}
    \frac{\partial (\Delta\psi)}{\partial t}=\mathcal{L}(\psi_0)\Delta\psi,
    \label{eq_3}
\end{equation}

\noindent where $\Delta\psi$ represents a small perturbation and $\mathcal{L}(\psi_0)$ is the corresponding linearized operator. The eigenvalues of $\mathcal{L}(\psi_0)$ define the dynamical spectrum and describe the time evolution of the waveform’s response to perturbations of $\psi_0$. The solution of Eq.~\ref{eq_3} over a small time step $\Delta t$ can be written as

\begin{equation}
    \Delta\psi(t_0+\Delta t)=e^{\mathcal{L}(\psi_0)\Delta t}\Delta\psi(t_0).
    \label{eq_4}
\end{equation}

Since any perturbation can be decomposed into a linear combination of eigenvectors $v_n$ and their corresponding eigenvalues $\lambda_n$, Eq.~\ref{eq_4} may be expressed as

\begin{equation}
    \Delta\psi(t_0+\Delta t)=\sum_n e^{\lambda_n\Delta t}a_n v_n,
    \label{eq_5}
\end{equation}

\noindent where $a_n$ are constant values. Equation~\ref{eq_5} shows that a perturbation grows exponentially when $\mathrm{Re}(\lambda_n)>0$, remains in the system as it is when $\mathrm{Re}(\lambda_n)=0$, or decays exponentially when $\mathrm{Re}(\lambda_n)<0$.
  
Among these modes, one eigenvalue is associated with the continuous translational symmetry of the soliton in the azimuthal coordinate, which we refer to as the position-shifting eigenvalue, $\lambda_{\text{ps}}$ \cite{Moille:2025}. In the absence of any trapping mechanism, this mode is neutrally stable ($\lambda_{\text{ps}}=0$) and is associated with the eigenvector $\partial \psi/\partial\theta$. All remaining eigenvalues typically have negative real parts, so perturbations along any other degree of freedom decay except for the soliton position, which results in the DKS performing a random walk. This effect can be incorporated into the LLE through an additional noise-driven linear drift, yielding

\begin{equation}
    \frac{\partial \psi}{\partial t} = -(1+i\alpha)\psi - \eta\Delta n(t) \frac{\partial \psi}{\partial \theta} - i \frac{\beta_2}{2} \frac{\partial^2 \psi}{\partial \theta^2} + i |\psi|^2 \psi + F_0,
    \label{eq_6}
\end{equation}

\noindent where $\Delta n(t)$ denotes the noise and $\eta = \partial \omega_{\text{rep}} / \partial \Delta n$ is a loss-normalized quantity that quantifies the variation in repetition-rate due to noise. The noise is determined using a Langevin equation, $d\Delta n/dt = -\Gamma_\mathrm{n} \Delta n + \zeta_\mathrm{n}(t)$, where $\zeta_\mathrm{n}$ is a white noise source and $\Gamma_\mathrm{n}$ is the dissipation rate at which the noise source relaxes in the system. The white noise source is defined as $\langle \zeta_\mathrm{n}(t)\zeta_\mathrm{n}(t-\tau)\rangle = A\,\delta(\tau)$, where $A$ is the variance of the noise source. This formulation is a general description of all noise sources that affect the repetition-rate of a DKS in microresonators, including thermorefractive noise (TRN), thermo-optic noise, and laser noise converted into repetition-rate noise via DW recoil and Raman SSFS. In Section 3, we model TRN specifically as a representative case.


When the input pump is phase modulated at a frequency of the FSR ($\zeta_1$), the soliton solution remains stationary relative to the frame of reference. For such a system, phase modulation of the pump can be accounted for in the LLE as \cite{Taheri:2015}

\begin{equation}
    \frac{\partial \psi}{\partial t} = -(1+i\alpha)\psi - \eta\Delta n(t) \frac{\partial \psi}{\partial \theta} - i \frac{\beta_2}{2} \frac{\partial^2 \psi}{\partial \theta^2} + i |\psi|^2 \psi + F_0 e^{i \delta_M \sin \theta},
    \label{eq_7}
\end{equation}

\noindent where $\delta_M$ is the modulation depth. 


In our proposed feedback controlled self-trapping system, the pump's phase is modulated by the RF signal that contains the intrinsic noise from the microresonator. Due to the delay in the feedback, the modulation frequency is a delayed version of the soliton repetition-rate, given by $\Delta \omega_{\mathrm{rep}}(t-t_{\mathrm{f}})$, where $t_{\text{f}}$ represents the feedback delay. The modified normalized LLE for this system becomes

\begin{equation}
    \frac{\partial \psi}{\partial t} = -(1+i\alpha)\psi + -\left[\Delta \omega_{\mathrm{rep}}(t-t_\mathrm{f})-\eta\Delta n(t)\right] \frac{\partial \psi}{\partial \theta} - i \frac{\beta_2}{2} \frac{\partial^2 \psi}{\partial \theta^2} + i |\psi|^2 \psi + F_0 e^{[i \delta_M \sin \theta]}.
    \label{eq_8}
\end{equation}

\noindent The principal difference between the external-trapping and self-trapping model is: in the self-trapping scheme, the modulation frequency is noisy. And this noise, $\Delta\omega_{\text{rep}}(t-t_{\text{f}})$, is correlated to the intrinsic noise of the microresonator.

\begin{figure}[!tb]
    \centering
    \includegraphics[width=1.0\linewidth]{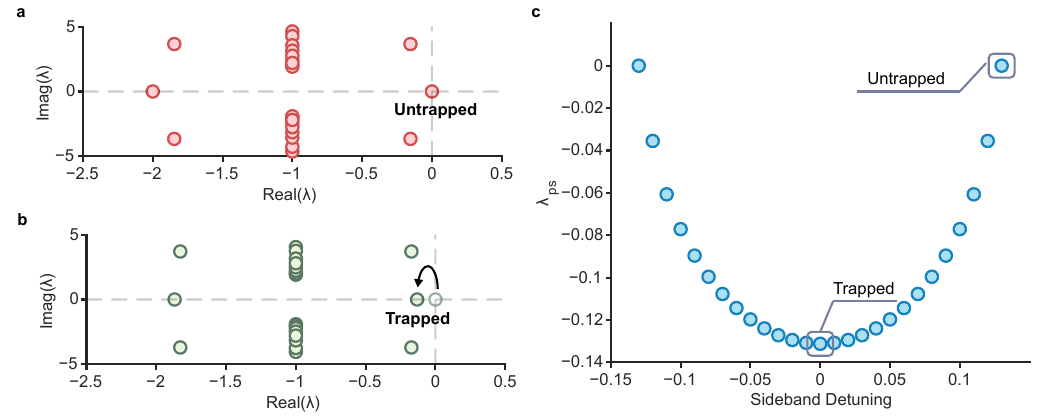}
    \caption{Dynamical spectrum corresponding to the (a) untrapped system and (c) a trapped system. (c) Change in position shifting eigenvalue as a function of sideband detuning, which is the deviation in modulation frequency from the FSR.}
    \label{fig:perturb&dynspec}
\end{figure}

To obtain repetition-rate noise power spectral density (PSD) of repetition-rate noise for the three types of systems discussed above, first, the LLE is solved using the split-step Fourier method \cite{Agrawal:2019}. The modulation depth is set to $\delta_M=0$ for the untrapped system and $\delta_M=0.65$ for the externally and feedback-controlled trapped systems, based on practical device parameters and similar values used in the literature \cite{Taheri:2015}. After each photon lifetime, the soliton's position is tracked by calculating its centroid ($\theta_0$) inside the cavity. The trajectory of the DKS is utilized to calculate the repetition rate noise as

\begin{equation}
    \Delta \omega_{\mathrm{rep}}^{(k)} = \frac{\theta_0^{(k)} - \theta_0^{(k-1)}}{\Delta t}\times R,
    \label{eq_9}
\end{equation}

\noindent where $k$ denotes the $k^{\text{th}}$ time step and $R$ is the radius of the microresonator cavity. The mean of $\Delta \omega_{\mathrm{rep}}$ is first subtracted to remove the DC offset, and a Blackman window \cite{blackman:1959} is applied to suppress spectral leakage. The one-sided PSD of the repetition-rate noise is then obtained as

\begin{equation}
    S_{\omega_{\mathrm{rep}}}(f) = \frac{2\left| \mathcal{F}\left\{ \Delta\omega_{\mathrm{rep}} \right\} \right|^2}{N f_s},
    \label{eq_10}
\end{equation}

\noindent where $\mathcal{F}$ denotes the fast Fourier transform, $N$ is the number of samples in the trajectory data, and $f_s = 1/\Delta t$ is the sampling frequency. The leading factor of 2 accounts for folding the negative-frequency content of the two-sided spectrum onto the positive-frequency axis, so that the total noise power is preserved in the one-sided representation.

We also evaluate the PSD of repetition-rate noise using analytical models to compare with our numerical results. For a system with no trapping present, the one-sided power spectral density of the repetition rate noise can be derived from a Langevin equation given as \cite{Moille:2025}

\begin{equation}
    S_{\Delta \omega_{\mathrm{rep}}}(\omega) = \frac{2\eta^2 A}{\Gamma_{\mathrm{n}}^2+\omega^2},
    \label{eq_11}
\end{equation}

\noindent where $\Gamma_\mathrm{n}$ is the dissipation rate, $\eta$ is  the variation in repetition rate with variation in noise source amplitude, $A$ is the noise amplitude, and $\omega$ is the offset frequency. 

The one-sided power spectral density of $\Delta \omega_{\mathrm{rep}}$ for the externally trapped system can be expressed as

\begin{equation}       
        S_{\Delta \omega_{\mathrm{rep}}}(\omega) = \frac{2\eta^2\omega^2 A}{(\kappa^2\lambda_{\mathrm{ps}}^2+\omega^2)(\Gamma_\mathrm{n}^2+\omega^2)}.
        \label{eq_12}           
\end{equation} 

When the pump phase-modulation is controlled using feedback, $\Delta \omega_{\mathrm{rep}}(t)$ becomes directly proportional to the delayed version of itself, $\Delta \omega_{\mathrm{rep}}(t-t_{\text{f}})$, where $t_\mathrm{f}$ is the feedback delay. We evaluate the one-sided power spectral density of $\Delta \omega_{\mathrm{rep}}$ for the self-trapped system as

\begin{equation}
    S_{\Delta \omega_{\mathrm{rep}}}(\omega) = \frac{2\omega^2 \eta^2 A}{(\Gamma_\mathrm{n}^2+\omega^2)[(\kappa \lambda_{\mathrm{ps}}+\omega \sin \omega t_{\mathrm{f}})^2+\omega^2(1-\cos{\omega t_{\mathrm{f}}})^2]}.
    \label{eq_13}
\end{equation}

\noindent Eqs.~\ref{eq_11},~\ref{eq_12}, and~\ref{eq_13} are used to predict the $f_{\text{rep}}$ noise spectra for untrapped, externally trapped, and self-trapped systems, respectively, which we then compare to the LLE simulations using Eqs.~\ref{eq_7} and~\ref{eq_8}.

\begin{figure}[!tb]
    \centering
    \includegraphics[width=1.0\linewidth]{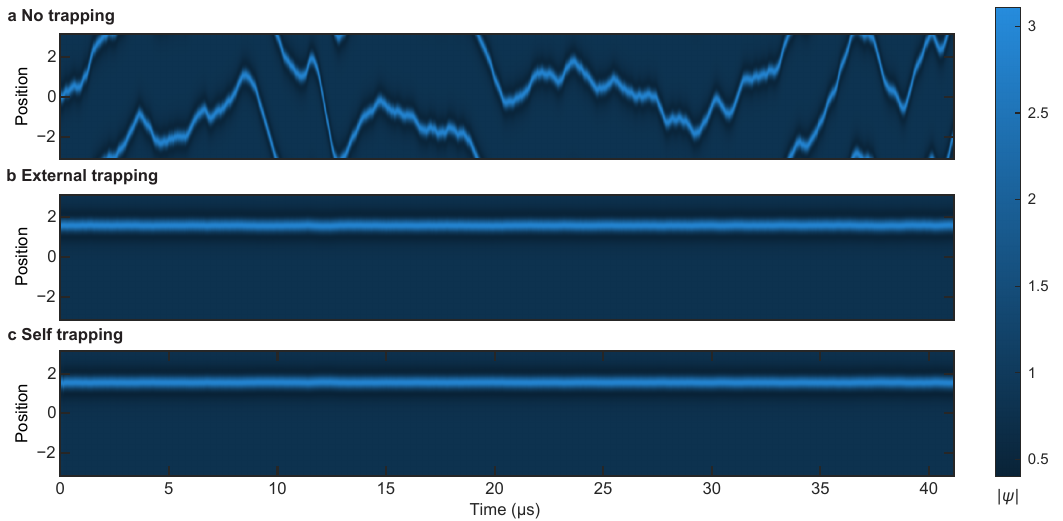}
    \caption{Numerical simulation of soliton evolution in the presence of $f_{\text{rep}}$ noise. When untrapped, the soliton does a random walk in the azimuthal domain (a). By contrast, the soliton does not visibly move when there is either (b) external or (c) feedback-controlled trapping}.
    \label{fig:time_evo}
\end{figure}

\section{Results and Discussion}\label{sec:results}
First, we examine the stability of the DKS in untrapped and trapped systems and its dependence on sideband detuning. The dynamical spectrum for the untrapped system is shown in Fig.~\ref{fig:perturb&dynspec}(a) where $\lambda_{\text{ps}} = 0$. In this regime, any perturbation that perturbs the soliton position persists, and the DKS position performs a noise-induced random walk. When the DKS is trapped via the pump's phase modulation, $\lambda_{\text{ps}}$ becomes negative, as shown in Fig.~\ref{fig:perturb&dynspec}(b), and perturbations to the soliton position decay exponentially. Furthermore, since trapping arises from sideband generation at the FSR frequency, any deviation of the modulation frequency from the FSR degrades the trapping strength. This effect is shown in Fig.~\ref{fig:perturb&dynspec}(c): as the deviation increases, $\lambda_{\text{ps}}$ returns to zero and the system becomes untrapped.

\begin{figure}[!tb]
    \centering
    \includegraphics[width=1.0\linewidth]{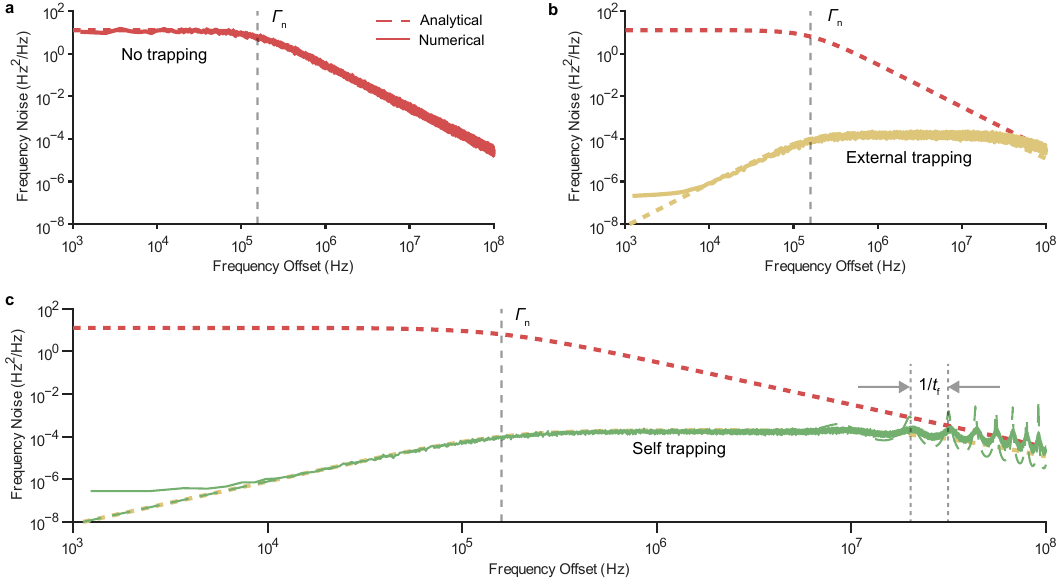}
    \caption{Theoretically and numerically obtained power spectral density of $f_{\text{rep}}$ noise for (a) unmodulated pumped system with no trapping, (b) externally trapped soliton system, and (c) the proposed feedback controlled self-trapping scheme.}
    \label{fig:white_noise}
\end{figure}

Next, we analyze the time evolution of the DKS for the three different systems. In Fig.~3, we show the time evolution of a single DKS, obtained from solving Eqs.~15 and 16.  In Fig.~3(a), we observe that the untrapped soliton undergoes a random walk.  By contrast, in Figs.~3(b) and 3(c), where the soliton is externally trapped or self-trapped respectively, no motion is observed.

\begin{figure}[!tb]
    \centering
    \includegraphics[width=1.0\linewidth]{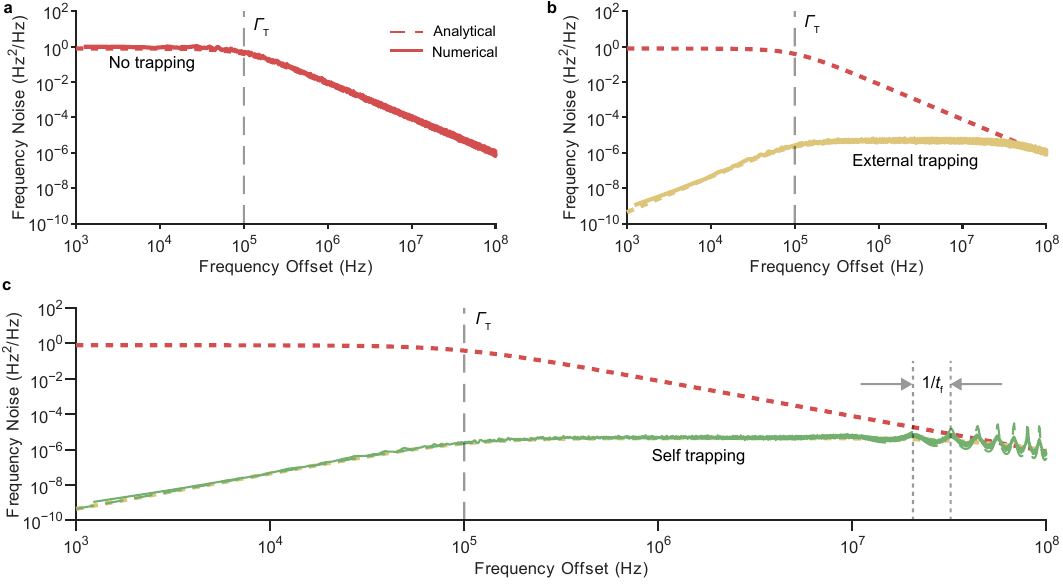}
    \caption{Theoretical and numerical power spectral density of $f_{\text{rep}}$ noise for (a) untrapped, (b) externally trapped, and (c) self-trapped systems when the noise source is only TRN instrinsic in the microresonator.}
    \label{fig:TRN}
\end{figure}
To compare the performance of the untrapped, externally trapped, and self-trapped systems, we numerically calculate the noise PSD for each configuration using the methodology described in Section 2. In our general noise model, we set $\Gamma_{\mathrm{n}} = 10^6\ \mathrm{Hz}$, a value on the order of, but not identical to, the thermal dissipation rate of $\mathrm{Si_3N_4}$. In practice, $\Gamma_{\mathrm{n}}$ can be adjusted to match the relevant noise source. For simplicity, we set $\eta = 1$ across all three configurations studied in this work. The variance of the stochastic noise source is chosen such that the resulting soliton-position jitter is clearly distinguished in our simulations. A total number of 120 simulations have been performed and averaged over for each setup to obtain accurate results in a Monte-Carlo fashion. The noise PSDs for general noise source are presented in Fig.~\ref{fig:white_noise}. It is observed from Fig.~\ref{fig:white_noise}(a) that an untrapped DKS drifts in azimuthal coordinates in response to the randomly varying noise source resulting in high noise in the frequency band of interest . When the DKS is  externally trapped using a very low-noise RF source, we observe a noise reduction of up to five orders of magnitude as shown in Fig.~\ref{fig:white_noise}(b) for frequencies lower than the cavity decay rate \cite{Weng:2019}. With the proposed feedback-controlled self-trapping, we achieve a noise suppression similar to that of the externally trapped system (Fig.~\ref{fig:white_noise}(c)). The oscillations at high offset frequencies are associated with the delay in feedback which is verified from the frequency difference between two adjacent oscillation peaks being exactly equal to $1/t_{\text{f}}$ and shown in Fig.~\ref{fig:white_noise}(c)). Despite the oscillations, the proposed system is still competitive for noise reduction in the frequency band of interest for many applications as well as being integration compatible. In addition, we use the derived theoretical models in Eqs.~\ref{eq_11}, \ref{eq_12}, and~\ref{eq_13} to calculate the PSDs of the three systems, and the results are illustrated in Fig.~\ref{fig:white_noise} using dashed lines. It is observed from Fig.~\ref{fig:white_noise} that the theoretical calculations agree well with our numerical results validating the performance of the proposed system.

Finally, we consider the case where the performance of the system is limited by TRN in the microresonator. For that, we compare the noise PSDs of the untrapped, externally trapped, and self-trapped systems when only TRN is present. TRN is critical in small mode-volume Si\textsubscript3N\textsubscript4 microresonators \cite{Drake:2020,Huang:2019}. Modeling the TRN is slightly different than our generalized model described earlier. For TRN modeling, we have followed the LLE based numerical framework described in \cite{Moille:2025}, where the pump detuning and dispersion parameter of the cavity are modified according to thermal fluctuations. The numerically obtained repetition-rate noise PSDs for our three configurations when only TRN is the source, are shown in Fig.~\ref{fig:TRN}.  It is observed from Fig.~\ref{fig:TRN}(a) that similar to the generalized white noise simulations, noise in the untrapped system is suppressed only for frequencies lower than the thermal dissipation rate, $\Gamma_T$. When the pump's phase is modulated using a low-noise external reference, noise is significantly decreased for frequencies above $\Gamma_T$ up to the cavity decay rate as shown in Fig.~\ref{fig:TRN}(b). Feedback-controlled self-trapping exhibits similar noise reduction as the externally trapped system (Fig.~\ref{fig:TRN}(c)). The results obtained using our derived theoretical models (dashed lines) match our numerical findings. 

\section{Conclusion}
In summary, we have proposed a feedback-controlled self-trapping configuration for microresonator based frequency combs that suppresses repetition-rate noise without requiring bulky RF generators. Using LLE-based Monte Carlo simulations, we compared the noise PSD of the previously reported external-trapping scheme with the proposed self-trapping scheme under a common noise source, and found that the noise suppression achieved by self-trapping is competitive with that of external trapping. We further analyzed the stability of the proposed scheme using dynamical methods and, from the resulting stability parameters, derived analytical models that predict the noise PSD of both configurations. The close agreement between these analytical predictions and our numerical simulations validates the accuracy of our analysis. The proposed configuration is well suited to integration in modern chip-scale photonic devices for frequency-comb stabilization where the final limit for the noise performance is set only by the pump.
\bibliographystyle{osajnl}
\bibliography{references}

\end{document}


\maketitle

\section{Derivation of Numerical Models} 

\setcounter{equation}{0}
\renewcommand{\theequation}{S.\arabic{equation}}

\subsection{No trapping}
The Lugiato--Lefever equation (LLE) \cite{Chembo:2013} is expressed as

\begin{equation}
    \frac{\partial \mathcal{A}}{\partial t}= -\left(\frac{\kappa}{2} + i\sigma\right)\mathcal{A}-\frac{\zeta_2}{2}\frac{\partial^2\mathcal{A}}{\partial\theta^2}-ig_0|\mathcal{A}|^2\mathcal{A}+\frac{\kappa}{2}\mathcal{F},
    \label{eq_1}
\end{equation}

\noindent where $\mathcal{A}$ denotes the slowly varying intracavity field envelope, $\kappa$ is the cavity decay rate, $\sigma$ is the detuning between the pump laser and the nearest cavity resonance, $\zeta_2$ is the second-order dispersion coefficient, $g_0$ is the four-wave mixing gain, and $\mathcal{F}$ is the pump amplitude. For convenience, Eq.~\ref{eq_1} is typically normalized by the cavity loss rate and rewritten in dimensionless form as

\begin{equation}
    \frac{\partial \psi}{\partial\tau}=-(1+i\alpha)\psi-i\frac{\beta}{2}\frac{\partial^2\psi}{\partial\theta^2}+i|\psi|^2\psi+F_0,
    \label{eq_2}
\end{equation}

\noindent where $\psi=\mathcal{A}/\mathcal{A}^*_{\mathrm{th}}$, $|\mathcal{A}_{\mathrm{th}}|=(\Delta\omega_0/2g_0)^{1/2}$, $F_0=\mathcal{F}/|\mathcal{A}^*_{\mathrm{th}}|$, $\tau$ is the normalized time, $\beta$ is the normalized second-order dispersion parameter, and $\alpha$ is the normalized detuning. Small perturbations in the system decay in all degrees of freedom except for the soliton position, leading to uncontrolled timing jitter around the cavity. This effect can be incorporated into the LLE through an additional noise-driven linear drift, yielding

\begin{equation}
\frac{\partial \psi}{\partial \tau} = -(1+i\alpha)\psi - \eta\Delta n(\tau) \frac{\partial \psi}{\partial \theta} - i \frac{\beta}{2} \frac{\partial^2 \psi}{\partial \theta^2} + i |\psi|^2 \psi + F_0,
\label{eq_3}
\end{equation}

\noindent where $\Delta n(\tau)$ denotes the stochastic noise source and $\eta = (\kappa/2) \partial \Delta \omega_{\mathrm{rep}}/\partial t$ quantifies the variation in the repetition rate ($f_{\text{rep}}$).


\subsection{External trapping}

Phase-modulating the pump with an external RF source generates equidistant sidebands around the pump frequency. The power of each sideband can be expressed mathematically as \cite{Taheri:2015}

\begin{equation}
    \mathcal{F}_m = \mathcal{F}_0 \mathcal{J}_m(\delta_M) e^{i[(\omega_{p} + m \omega_M) - (\omega_0 + \omega_m)]t},
    \label{eq_4}
\end{equation}

\noindent where $\mathcal{J}_m$ is the Bessel function of the first kind representing the amplitude of the $m$-th sideband, $\delta_M$ is the modulation depth, $\omega_p$ is the pump frequency, $\omega_M$ is the modulation frequency, $\omega_0$ is the central resonance frequency, and $\omega_m = m \zeta_1 + D_{\mathrm{int}}$, where $D_{\mathrm{int}}$ is the integrated dispersion. Hence, the driving power for the dissipative Kerr soliton (DKS) is written as 

\begin{equation}
    \mathcal{F} = \sum_{m = -\infty}^{\infty} \mathcal{F}_0 \mathcal{J}_m(\delta_M) e^{i[(\sigma + m \omega_M + D_{\mathrm{int}}) - (m\zeta_1 + D_{\mathrm{int}}) ]t}.
    \label{eq_5}
\end{equation}

\noindent Using the Jacobi-Anger expansion, Eq.~\ref{eq_5} can be rewritten as \cite{Taheri:2015}

\begin{equation}
    \mathcal{F} = \mathcal{F}_0 e^{i\sigma t} \sum_{m = -\infty}^{\infty}  \mathcal{J}_m(\delta_M) e^{i[m \omega_M - m\zeta_1]t} = \mathcal{F}_0 e^{i\sigma t} e^{i\delta_M \sin((\omega_M-\zeta_1)t)}.
    \label{eq_6}
\end{equation}

\noindent Substituting Eq.~\ref{eq_6} into Eq.~\ref{eq_1}, we obtain

\begin{equation}
    \frac{\partial \mathcal{A}}{\partial t} = -\frac{\kappa}{2} \mathcal{A} - \frac{\zeta_2}{2} \frac{\partial^2 \mathcal{A}}{\partial \theta^2} - i g_0 | \mathcal{A} |^2 \mathcal{A} + \frac{\kappa}{2} \mathcal{F}_0 e^{i\sigma t} e^{i\delta_M \sin((\omega_M-\zeta_1)t)}.
    \label{eq_7}
\end{equation}

\noindent To make \ref{eq_7} autonomous, we shift to a rotating frame of reference where $(\omega_M-\zeta_1)t=\theta'$ and $t=t'$. Consequently, the derivatives transform as $\partial/\partial\theta = \partial/\partial\theta'$ and $\partial/\partial t = -(\omega_M-\zeta_1)\partial/\partial\theta'+\partial/\partial t'$. Applying these transformations to Eq.~\ref{eq_7}, substituting $\mathcal{A} \rightarrow \mathcal{A}e^{i\sigma t}$, and dropping the primes for convenience, we obtain

\begin{equation}
    \frac{\partial \mathcal{A}}{\partial t} = -\left( \frac{\kappa}{2} + i \sigma \right) \mathcal{A} + \left( \omega_M - \zeta_1 \right) \frac{\partial \mathcal{A}}{\partial \theta} - \frac{\zeta_2}{2} \frac{\partial^2 \mathcal{A}}{\partial \theta^2} - i g_0 | \mathcal{A} |^2 \mathcal{A} + \frac{\kappa}{2} \mathcal{F}_0 e^{i\delta_M \sin \theta}.
    \label{eq_8}
\end{equation}

\noindent Normalization of this modified LLE with respect to $\kappa$ yields

\begin{equation}
    \frac{\partial \psi}{\partial \tau} = -(1+i\alpha)\psi + (\omega_M-\zeta_1) \frac{\partial \psi}{\partial \theta} - i \frac{\beta}{2} \frac{\partial^2 \psi}{\partial \theta^2} + i |\psi|^2 \psi + F_0 e^{i \delta_M \sin \theta}.
    \label{eq_9}
\end{equation}

\noindent If the pump is phase-modulated at the FSR frequency, $\omega_M = \zeta_1$, hence, the DKS remains stationary in this reference frame. Incorporating the microresonator's intrinsic noise, $\omega_M - (\zeta_1+\eta\Delta n)=-\eta\Delta n$. Hence, the LLE becomes

\begin{equation}
    \frac{\partial \psi}{\partial \tau} = -(1+i\alpha)\psi - \eta\Delta n(\tau) \frac{\partial \psi}{\partial \theta} - i \frac{\beta}{2} \frac{\partial^2 \psi}{\partial \theta^2} + i |\psi|^2 \psi + F_0 e^{i \delta_M \sin \theta}.
    \label{eq_10}
\end{equation}

\subsection{Feedback controlled self-trapping}
In our proposed feedback-controlled self-trapping scheme, the pump phase is modulated by an RF signal that includes the microresonator's intrinsic noise. Due to the feedback loop's delay ($t_{\text{f}}$), the modulation frequency acts as a delayed version of the soliton repetition rate, $\Delta \omega_{\mathrm{rep}}(\tau-\tau_{\mathrm{f}})$. The modified LLE may be written as

\begin{equation}
\begin{split}
    \frac{\partial \mathcal{A}}{\partial t} = -\left( \frac{\kappa}{2} + i \sigma \right) \mathcal{A} + \left[ \zeta_1+\Delta \omega_{\mathrm{rep}}(\tau-\tau_{\mathrm{f}}) - \zeta_1 - \eta\Delta n(\tau) \right] \frac{\partial \mathcal{A}}{\partial \theta} \\ 
    - i\frac{\zeta_2}{2} \frac{\partial^2 \mathcal{A}}{\partial \theta^2} - i g_0 | \mathcal{A} |^2 \mathcal{A} + \frac{\kappa}{2} \mathcal{F}_0 e^{i\delta_M \sin \theta}.
    \label{eq_11}
\end{split}
\end{equation}

\noindent Normalizing Eq.~\ref{eq_11} for cavity loss yields 

\begin{equation}
    \frac{\partial \psi}{\partial \tau} = -(1+i\alpha)\psi + \left[\Delta \omega_{\mathrm{rep}}(\tau-\tau_\mathrm{f})-\eta\Delta n(\tau) \right] \frac{\partial \psi}{\partial \theta} - i \frac{\beta}{2} \frac{\partial^2 \psi}{\partial \theta^2} + i |\psi|^2 \psi + F_0 e^{i \delta_M \sin \theta}.
    \label{eq_12}
\end{equation}

\section{Derivation of Analytical Models}

\subsection{No trapping}

\noindent Because the DKS evolution is driven by both deterministic and random forces, we model it using the Langevin equation

\begin{equation}
    \frac{d\Delta n}{dt} = -\Gamma_\mathrm{n} \Delta n + \zeta_\mathrm{n}(t),
    \label{eq_13}
\end{equation}

\noindent where $\Gamma_\mathrm{n}$ is the noise dissipation rate for a general repetition-rate noise source, $\Delta n$ is the noise amplitude, and $\zeta_{\mathrm{n}}(t)$ is a white noise source with the autocorrelation function 

\begin{equation}
    \langle \zeta_\mathrm{n}(t)\zeta_\mathrm{n}(t-\tau)\rangle = A \delta(\tau),  
    \label{eq_14}
\end{equation}

\noindent where $A$ is a constant. Taking the Fourier transform of Eq.~\ref{eq_13} yields

\begin{align}
    \begin{split}
        -i\omega \Delta \Tilde{n}(\omega) &= -\Gamma_\mathrm{n} \Delta \Tilde{n}(\omega) + W(\omega) \\
        \Rightarrow \Delta \Tilde{n}(\omega) &= \frac{W(\omega)}{\Gamma_{\mathrm{n}} - i \omega},
    \end{split}
    \label{eq_15}
\end{align}

\noindent where $W(\omega)$ is the Fourier transform of the noise autocorrelation function. Without trapping, the repetition-rate noise is directly proportional to the source fluctuations and thus can be written as

\begin{equation}
    \Delta \omega_{\mathrm{rep}} = \eta \times \Delta n,
    \label{eq_16}
\end{equation}

\noindent where $\eta = \partial \omega_{\mathrm{rep}}/\partial n$ characterizes how the repetition rate responds to the noise source. Taking the time derivative of  Eq.~\ref{eq_16} yields

 \begin{align}
        \begin{split}
            \frac{d \Delta \omega_{\mathrm{rep}}}{dt} &= \eta \frac{d\Delta n}{dt} \\
            \Rightarrow \Delta \Tilde{\omega}_{\mathrm{rep}}(\omega) &=  \frac{\eta W(\omega)} {\Gamma_{\mathrm{n}} - i\omega}. \\  
        \end{split}       
        \label{eq_17}
    \end{align}    
    
\noindent Consequently, the one-sided power spectral density (PSD) of the repetition-rate noise is  

\begin{equation}
    S_{\omega_{\mathrm{rep}}}(\omega) = \langle  \left| \Delta \Tilde{\omega}_{\mathrm{rep}} \right|^2 \rangle = \eta^2 \left< \frac{\left| W(\omega) \right|^2}{\left| \Gamma_{\mathrm{n}} - i\omega \right|^2} \right> = \frac{2\eta^2 A}{\Gamma_{\mathrm{n}}^2+\omega^2}.
    \label{eq_18}
\end{equation}

\subsection{External trapping}
\noindent Phase modulation induces a trapping potential that causes small perturbations to decay exponentially. The corresponding Langevin equation can be expressed as

\begin{equation}
   \frac{d \Delta \omega_{\mathrm{rep}}(t)}{dt} = \kappa\lambda_{\mathrm{ps}} \Delta \omega_{\mathrm{rep}}(t) + \eta \frac{d \Delta n}{dt},
   \label{eq_19}
\end{equation}

\noindent where $\kappa$ is the cavity decay rate (the reciprocal of the photon lifetime, $\tau_{\text{p}}$) and $\lambda_{\mathrm{ps}}$ is the position-shifting eigenvalue. The one-sided PSD of $\Delta \omega_{\mathrm{rep}}$ for this externally trapped system is

\begin{equation}        
        S_{\omega_{\mathrm{rep}}}(\omega) = \frac{2\eta^2\omega^2 A}{(\kappa^2\lambda_{\mathrm{ps}}^2+\omega^2)(\Gamma_\mathrm{n}^2+\omega^2)}.
        \label{eq_20}            
\end{equation} 

\subsection{Self trapping}
\noindent In the feedback-controlled scheme, the repetition-rate noise $\Delta \omega_{\mathrm{rep}}(t)$ is coupled to its delayed version, $\Delta \omega_{\mathrm{rep}}(t-t_{\text{f}})$. The governing Langevin equation can be written as

\begin{align}
    \begin{split}
        \frac{d \Delta \omega_{\mathrm{rep}}(t)}{dt} &= \kappa\lambda_{\mathrm{ps}} \Delta \omega_{\mathrm{rep}}(t) + \eta  \frac{d \Delta n(t)}{dt} + \frac{d \Delta \omega_{\mathrm{rep}}(t-t_\mathrm{f})}{dt}. 
        \label{eq_21}
    \end{split}        
\end{align}

\noindent By taking the Fourier transform of Eq.~\ref{eq_21}, we obtain

\begin{equation}    
        \Delta \Tilde{\omega}_{\mathrm{rep}}(\omega) =  \frac{i\omega\eta  W(\omega)}{(\Gamma_{\mathrm{n}} - i\omega)[\kappa\lambda_{\mathrm{ps}} + i\omega(1-e^{i\omega t_{\mathrm{f}}})]}.    
        \label{eq_22}    
\end{equation}

\noindent Using Eq.~\ref{eq_22}, the one-sided PSD of the repetition-rate noise for the feedback-controlled scheme is

\begin{equation}
    S_{\Delta \omega_{\mathrm{rep}}}(\omega) = \frac{2\omega^2 \eta^2 A}{(\Gamma_\mathrm{n}^2+\omega^2)[(\kappa \lambda_{\mathrm{ps}}+\omega \sin \omega t_{\mathrm{f}})^2+\omega^2(1-\cos{\omega t_{\mathrm{f}}})^2]}.
    \label{eq_23}
\end{equation}

\noindent Equations~\ref{eq_18}, \ref{eq_20}, and \ref{eq_23} predict the $f_{\text{rep}}$ noise spectra for the untrapped, externally trapped, and self-trapped systems, respectively. These analytical models provide a baseline to validate our numerical simulations. 

\section{Simulation Details}
\subsection{Generalized noise simulation setup}
Our simulations model a 100 GHz microresonator using the parameters summarized in Table~\ref{tab:1}. Noise variance, $A$, was slowly increased from small until we found an $A$ where the DKS jitter was measurable in the simulations. This value of $A$ is used throughout our generalized noise simulations. The dissipation rate $\Gamma_n$ is chosen as a value similar to the thermal dissipation; although, it can be varied depending on the noise source. Modulation strength, $\delta_M$ is chosen based on practical parameters and similar value used in the literature \cite{Taheri:2015}. 

\begin{table}[ht]
    \centering
        \begin{tabular}{l l l}
        
        \toprule 
        \textbf{Parameter} & \textbf{Value} \\
        \midrule 
        $Q$ & $10^6$ \\
        \hline
        $\kappa$ & $1.22\times10^9$ Hz  \\
        \hline
        $R$ & $225$ $\mu$m \\
        \hline
        $\alpha$ & $3.2$ \\
        \hline
        $\beta$ & $-0.2$ \\
        \hline
        $F$ & $2.2$ \\
        \hline
        $\Gamma_n$ & $10^6/2\pi$ rad/s \\
        \hline
        $\eta$ & $\kappa/2$ \\
        \hline
        $\delta_M$ & $0.65$ \\
        \hline
        $A$ & $10$ \\         
        \bottomrule 
        \end{tabular}
    \caption{\label{tab:1}Simulation parameters when a general white noise source is considered}
\end{table}

\begin{table}[ht]
    \centering
        \begin{tabular}{l l l}
        
        \toprule 
        \textbf{Parameter} & \textbf{Value} \\
        \midrule 
        $Q$ & $10^6$ \\
        \hline
        $\kappa$ & $1.22\times10^9$ Hz  \\
        \hline
        $R$ & $225$ $\mu$m \\
        \hline
        $\alpha$ & $3.2$ \\
        \hline
        $\beta$ & $-0.2$ \\
        \hline
        $F$ & $2.2$ \\
        \hline
        $\Gamma_T$ & $10^5$ Hz \\
        \hline
        $\eta_T$ & $4.55\times10^7$ \\
        \hline
        $\delta_M$ & $0.65$ \\
        \hline
        $k_B$ & $1.38\times10^{-23}$ J/K \\
        \hline
        $\rho$ & $3000 \; \rm{kg\cdot m^{-3}}$ \\
        \hline
        $C$ & $700\;\rm{J\cdot kg^{-1}K^{-1}}$ \\
        \hline
        $V$ & $800\;\rm{\mu m^3}$ \\
        \hline
        $m$ & $2000$ \\        
        \bottomrule 
        \end{tabular}
    \caption{\label{tab:2}Simulation parameters when only TRN is considered}
\end{table}

\subsection{Thermorefractive noise simulation setup}
We model thermorefractive noise (TRN) following the approach outlined in \cite{Moille:2025}. The autocorrelation function of the TRN source is given by \cite{Huang:2019}:

\begin{equation}
    \langle\zeta_T(t)\zeta_T(t+\tau)\rangle = \frac{\Gamma_T k_B T^2}{\rho C V},
\end{equation}

\noindent where $\Gamma_T$ is the thermal dissipation rate, $k_B$ is the Boltzmann constant, $T$ is the absolute temperature, $\rho$ is the material density, $C$ is the specific heat capacity, and $V$ is the modal volume of the fundamental TE mode. Table~\ref{tab:2} lists the parameters used for the TRN simulations. Majority of the parameters are taken from Ref. \cite{Moille:2025}, except $V$ and $m$ are calculated using FEM modeling.

\bibliographystyle{osajnl}
\bibliography{references}